\documentclass[a4paper,12pt]{article}

\usepackage{subfigure}
\usepackage{graphicx}
\usepackage{floatflt}
\usepackage{wrapfig}
\usepackage{cite}

\begin{document}

\title{ \textsc{scaling separability criterion: application to gaussian states}}

\author{\bf Alexandr Sergeevich$^{1*}$ and Vladimir I. Man'ko$^{2}$}
\date{}
\maketitle

\centerline{$^{1}$\textit{School of Physics, The University of
Sydney, New South Wales 2006, Australia}} \vspace{1mm}

\centerline{$^{2}$\textit{P. N. Lebedev Physical Institute, Russian
Academy of Sciences,}}

\centerline{\textit{Leninskii Prospect 53, Moscow 119991, Russia}}

\vspace{4mm} \centerline{*corresponding author e-mail:
a.sergeevich@physics.usyd.edu.au}

\centerline{e-mail: manko@sci.lebedev.ru}

\def\thesection{\arabic{section}.}
\makeatletter
\renewcommand\@biblabel[1]{#1.}
\makeatother

\vspace{5mm}
\begin{abstract}
\noindent We introduce examples of three- and four-mode entangled Gaussian mixed states that are not detected
by the scaling and Peres–Horodecki separability criteria. The presented modification of the scaling
criterion resolves this problem. Also it is shown that the new criterion reproduces the main features of
the scaling pictures for different cases of entangled states, while the previous versions lead to completely
different outcomes. This property of the presented scheme is evidence of its higher generality.
\end{abstract}


\setcounter{section}{0} \setcounter{equation}{0}

\section{Introduction}
Being one of the most striking phenomena in quantum physics, entanglement \cite{1,2,3} has been thoroughly
investigated for many years. One of the main research directions aims at finding the universal
separability criterion \cite{4,5,6,7,8,9}. The particular case of Gaussian states was investigated using the Peres-Horodecki separability criterion \cite{4,5}, which is applicable to two-mode states that fail to work in general
when applied to cases of higher dimensions. The Peres–Horodecki criterion is based on the so-called
ppt-transform (partial positive transpose transform), i.e., transpose of the density matrix of one part of
a multipartite state, leaving other states untouched. If after the ppt-transform the density matrix of the
whole system becomes physically meaningless, then the state is entangled. This is the essence of the
Peres–Horodecki separability criterion.

Our investigation concerns the previously developed extension of this criterion called "scaling separability
criterion" \cite{10,11,12}. In considering the density matrix transpose as a time reversal, we can think
of generalizing this operation to time scaling. Doing so, the scaling criterion detects more Gaussian
entangled states, yielding a nonphysical density matrix for at least one scaling parameter.

The separability criteria are closely connected with the question of finding the measure of entanglement
(i.e., \cite{13,14}), some of which are based on the Peres–Horodecki criterion. The scaling criterion provides an intuitive visual measure as well, being not operational though. The criterion itself finds
application in different cases discussed in \cite{15,16,17}.

In this paper, we provide a modification of the scaling criterion, which allows one to detect entanglement
in a wide range of Gaussian pure and mixed states. Based on the results of our previous paper \cite{12},
in this work we investigate the use of the scaling criterion of separability, looking at cases for which the
previous version of the criterion did not work as well as the Peres–Horodecki criterion. Demonstrating
the power of this method, we also discuss its higher generality, comparing the results for different related
cases of mixed Gaussian states.

The paper is organized as follows.

In Sec. 2, we provide the basic theoretical background on the scaling criterion of separability applied
to Gaussian states. Section 3 is devoted to the discussion of multimode uncertainty relations, which are
needed for detecting nonphysical density matrices, on which the criterion is based. Finally, in Sec. 4
we provide examples of the method of operation for some three- and four-mode Gaussian mixed states,
also discussing the aspects of the criterion, such as a comparison with its previous version and the
Peres–Horodecki criterion, and possible interpretation of the results, such as the entanglement measure,
etc.

\section{Scaling Criterion}

Let us consider a single mode photon state with the density matrix
$\rho$ which should obey the following conditions:

\begin{equation}\rho^+ = \rho,\;\;\;\mbox{Tr} \rho = 1,\;\;\;\rho\geq0. \label{rhobasic}\end{equation}

Also let $\hat q$ and $\hat p$ be the quadrature operators of this
state. Then we can rewrite relations (\ref{rhobasic}) in the
form of Robertson-Schr\"{o}dinger uncertainty relation
\cite{18,19}:

\begin{equation}\left(\begin{array}{cc}
\sigma_{q q} & \sigma_{q p}-i/2 \\
\sigma_{q p}+i/2 & \sigma_{p p} \\
\end{array} \right) \geq 0\end{equation}
where $\sigma_{\xi \zeta}=\left<\hat\xi\hat\zeta\right>$ and the
inequality is considered (here and further) in the sense of positivity
of all principal minors of the matrix. This condition can be easily
simplified to
\begin{equation}\Delta=\sigma_{q q}\sigma_{p p}-\sigma_{q p}^{2} \geq
\frac{1}{4} \label{unmodifieduncert1D}
\end{equation}

Now we'll transform the given state by multiplying its momentum by a
scaling parameter $\lambda$. This is equivalent to the
transform of time $t \rightarrow \lambda t$. Rewriting relation (\ref{unmodifieduncert1D}) for the modified state, we obtain the
following condition, which should hold if the new state is physically realizable:
\begin{equation}\frac{\Delta}{\lambda^2}=\frac{1}{\lambda^2}(\sigma_{q q}\sigma_{p p}-\sigma_{q p}^{2}) \geq
\frac{1}{4}
\end{equation}
Hence, the uncertainty relation holds for $\lambda\in
\left[-2\sqrt{\Delta},2\sqrt{\Delta}\right]$. It is worth mentioning
that the transform performed in the Peres-Horodecki criterion
($t\rightarrow-t$ or $\rho\rightarrow\rho^T$) is included in this
set of maps being represented by $\lambda=-1$ since $\Delta$ is
always greater than $\frac{1}{4}$. The scaling criterion of
separability considered in \cite{12} used this transform for
only $\lambda\in\left[-1,1\right]$, which, as it will be shown later,
is not enough for the entanglement detection.

The scaling criterion of separability for $n$-mode photon Gaussian
states tells us that if we'll apply the scaling transform with
coefficients $\lambda_i \in \left[-1,1\right]$ to the momenta $p_i$
of all the submodes, the modified state will become not physically
realizable (the Robertson-Schr\"{o}dinger uncertainty relations will
not hold) for some set $\{\lambda_i\}$ only if the state is
entangled. In \cite{12} we proved that for pure three- and
four- mode Gaussian states we can say "only if" and also that this
criterion is more powerful than the Peres-Horodecki criterion in
general. But there are some mixed entangled states that are not detected when using the scaling in
the $\left[-1,1\right]$ range. And further we will present some examples to show that these states can be detected
by the criterion if we choose $\lambda_i$ from the range
$\left[-2\sqrt{\Delta_i},2\sqrt{\Delta_i}\right]$, where $\Delta_i$
is the value of the left-hand side of the uncertainty relation
(\ref{unmodifieduncert1D}) for the $i$th submode.

\section{Multimode Uncertainty Relations}

There are many ways of checking if the state represented by some formula has a physical meaning. In our case, the simplest way is to check the fulfilment of the uncertainty relations in the Robertson–Schr\"{o}dinger form.

Let us introduce these relations in the multimode case, their modification under scaling transform,
and find the operations needed to apply them to the separability criterion.

The Wigner function of the generic Gaussian form for $n$-mode state reads
\begin{equation}W(q,p)=\frac{1}{\sqrt{\mathrm{det}\,
\sigma}}\exp\left(-\frac{1}{2} \bf Q \sigma^{-1} \bf Q ^T\right),
\end{equation}
where the $2n$-dimensional vector $\bf Q$ is
\begin{equation}\mathbf{Q}=\left(q_1-\left\langle q_1 \right\rangle, q_2-\left\langle
q_2 \right\rangle, ..., q_n-\left\langle q_n \right\rangle,
p_1-\left\langle p_1 \right\rangle, p_2-\left\langle p_2
\right\rangle, ..., p_n-\left\langle p_n \right\rangle \right),
\end{equation} and the matrix $\sigma$ is a $2n\times2n$ real
symmetric variance matrix
\begin{equation}\sigma_{r_i r_j}=\frac{1}{2}\left\langle \hat r_i \hat r_j + \hat r_j \hat r_i \right\rangle, \end{equation}
where $\hat r_i = \hat q_i$, $\hat r_{n+j}=\hat p_j$, and $i,j=1,...,n$.

The matrix $\sigma$ is a good characteristic of a Gaussian state by
which one can judge on its reality using the uncertainty relation
\begin{equation}\sigma+\frac{i}{2}\Omega \geq 0\end{equation}
where $\Omega$ presented in a block form is
$$\Omega=\left(\begin{array}{cc}
0 & -I \\
I & 0
\end{array}\right),$$ and $I$ is the identity matrix. Again, the
inequality here is considered as a nonnegativity of the principal minors
of the matrix.

It is easy to see that the scaling of the $i$th mode $p_i \rightarrow
\lambda_i p_i$ is identical to the division of $(n+i)$th row and
column of $\sigma$ by $\lambda_i$. Since the first $n$ principal
minors remain unchanged, the check of the uncertainty relations
should be performed for only minors of higher dimensions. Obviously,
one should perform the scaling in the range identified above and
check the positivity of each of the minors for all possible sets
$\{\lambda_i\}$. But further we will be considering only the determinant; to the best of our knowledge, there are
no entangled states that are detected by the lower minors and not detected by the determinant, and no
entangled Gaussian states that are not detected by the presented algorithm.

\section{Performance on Three- and Four-Mode Mixed States}

Let us provide some examples for which the scaling criterion (and
the Peres-Horodecki criterion) does
not work when using only scaling from $-1$ to $1$.

Performing the scaling of the three-mode mixed state with the
dispersion matrix
\begin{equation}\sigma = \left(\begin{array}{cccccc}
6/5 & 1/5 & 1/5 & 1/10 &  1/10 &  1/10 \\
1/5 & 6/5  & 1/5 &  1/10 &  1/10 &  1/10 \\ 
1/5 & 1/5 & 6/5 &  1/10 &  1/10 &  1/10 \\
1/10 &  1/10 & 1/10 & 1 & -1/8 & -1/8 \\  
1/10 &  1/10 &  1/10 & -1/8 & 1 & -1/8 \\  
1/10 &  1/10 &  1/10 & -1/8 & -1/8 & 1
\end{array}\right) \label{3mBadMatrix} \end{equation}
we'll get some area of negativity within our scaling range although
there is no negative results for $\lambda_i \in [-1,1]$. This result
is vividly depicted in Fig. 1 where the plot for
$\mathrm{det}\left(\sigma+\frac{i}{2}\Omega\right)$ versus $\lambda_2$ and $\lambda_3$ is shown.
Here $\lambda_1 = 1/2$, but other possible values of this coefficient will yield a very similar plot, also positive
on $\left[−1, 1\right]$. The grayscale depicts the level of negativity of the determinant (level of nonrealizability of
the state – how large is the violation of the uncertainty relations), and the white area stands for positive
values that we are not interested in. For comparison, Fig. 2 shows the plot for $\sigma$ with $1/5$ in the upper-right
and lower-left $4 \times 4$ quadrants. This is the one considered in our previous paper \cite{12}, and it is obvious
that there are negative points within $\lambda_2 \times \lambda_3 \in \left[−1, 1\right]$, while the first state is not detected as entangled either by the previous version of the scaling criterion or by the Peres–Horodecki criterion.

\begin{figure}[h]
  \begin{minipage}[t]{0.47\linewidth}
        \includegraphics[width=213 pt]{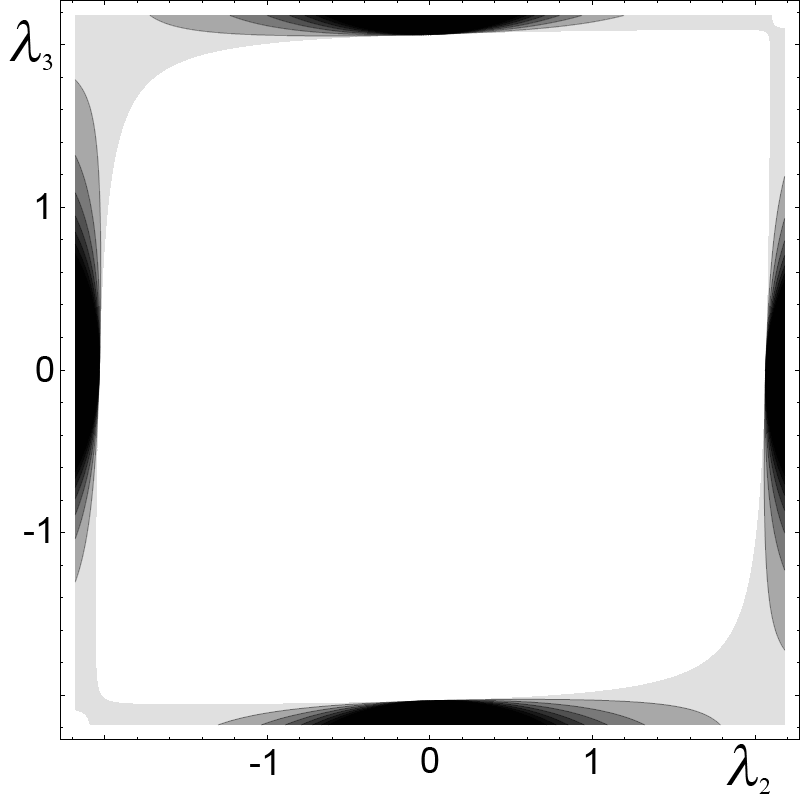}
\textbf{Fig. 1.} The separability test plot for the three-mode
state (9).
   \end{minipage}
   \begin{minipage}[t]{0.03\linewidth}
   $\;$
     \end{minipage}
 \begin{minipage}[t]{0.47\linewidth}
        \includegraphics[width=213 pt]{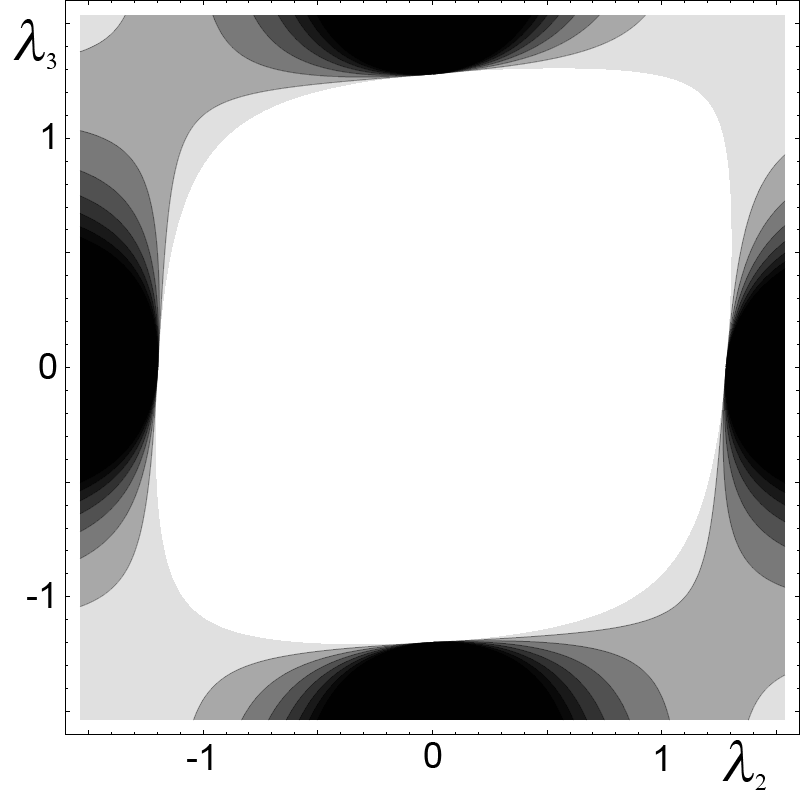}
\textbf{Fig. 2.} The separability test plot for the modified three-mode
state obtained from (9) by replacement of the matrix elements.
     \end{minipage}
\end{figure}

\begin{figure}[h]
  \begin{minipage}[t]{0.47\linewidth}
        \includegraphics[width=213 pt]{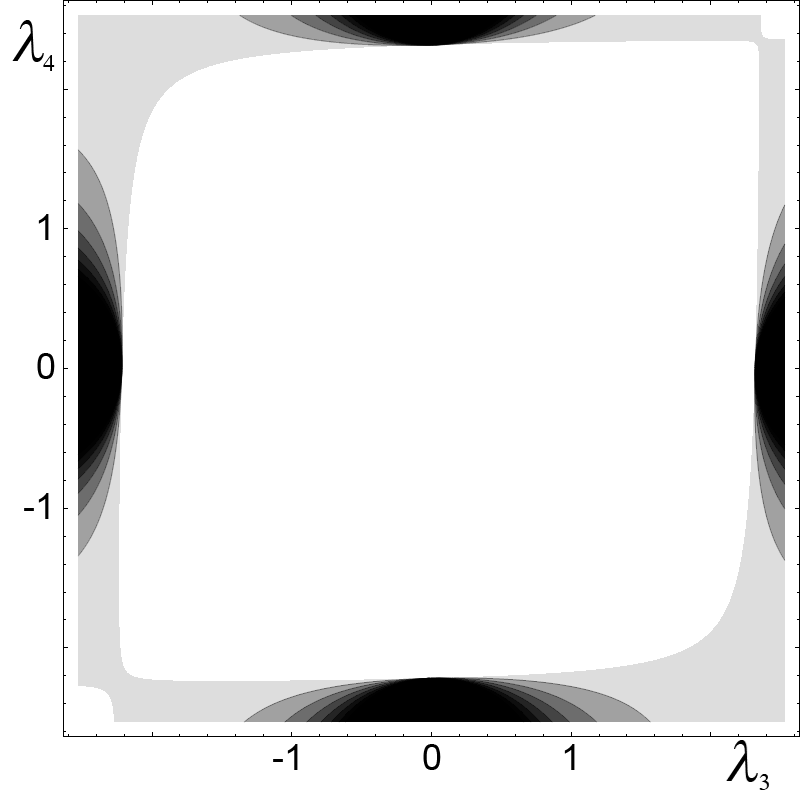}
\textbf{Fig. 3.} The separability test plot for the four-mode
state (10).
   \end{minipage}
      \begin{minipage}[t]{0.03\linewidth}
   $\;$
     \end{minipage}
 \begin{minipage}[t]{0.47\linewidth}
        \includegraphics[width=213 pt]{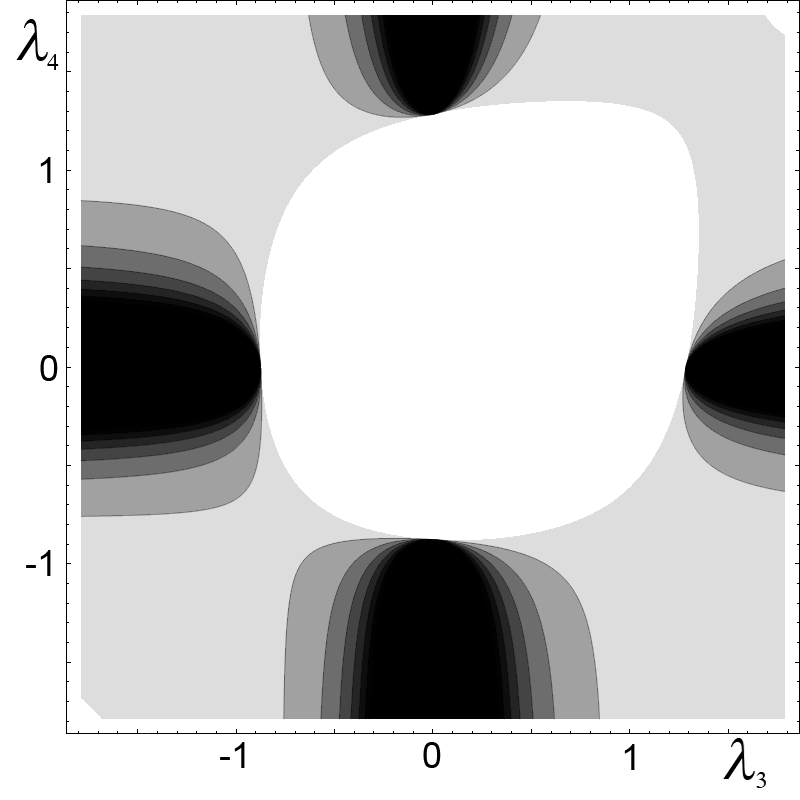}
\textbf{Fig. 4.} The separability test plot for the modified four-mode
state obtained from (10) by replacement of the matrix elements.
     \end{minipage}
\end{figure}

The same result is obtained for the following four-mode dispersion
matrix
\begin{equation}\sigma = \left(\begin{array}{cccccccc}
8/5 & 2/5 & 2/5 & 2/5 & 1/50 & 1/50 & 1/50 & 1/50 \\
2/5 & 8/5 & 2/5 & 2/5 & 1/50 & 1/50 & 1/50 & 1/50 \\
2/5 & 2/5 & 8/5 & 2/5 & 1/50 & 1/50 & 1/50 & 1/50 \\
2/5 & 2/5 & 2/5 & 8/5 & 1/50 & 1/50 & 1/50 & 1/50 \\
1/50 & 1/50 & 1/50 & 1/50 & 1 & -1/8 & -1/8 & -1/8 \\
1/50 & 1/50 & 1/50 & 1/50 & -1/8 & 1 & -1/8 & -1/8 \\
1/50 & 1/50 & 1/50 & 1/50 & -1/8 & -1/8 & 1 & -1/8 \\
1/50 & 1/50 & 1/50 & 1/50 & -1/8 & -1/8 & -1/8 & 1 \\
\end{array} \right)
\end{equation}
The plot for $\lambda_1 = 1$ and $\lambda_2 = \frac{1}{2}$ is shown
in Fig. 3 and also the previously considered state \cite{12} when the upper-right
and lower-left $6\times 6$ quadrants are filled with $1/10$
as shown on the Fig. 4. Again, the first state turns out to be realizable for all $\lambda_i \in \left[−1, 1\right]$, and so the entanglement would not be detected.

Now we see that this modified criterion reproduces the appearance of
the scaling picture for different states (for more examples, refer to
\cite{12}) and detects the entanglement for a larger class
of mixed states surpassing the power of its predecessors.

\section{Conclusions}
In this work, we showed that the new version of the scaling separability criterion is more powerful
than the initial one which, in turn, is stronger than the Peres–Horodecki criterion. Investigating some
examples of three- and four-mode mixed Gaussian states, we also provided an intuitive argument in favor
of the generality of this method, relating to the measure of entanglement as well.

\section*{Acknowledgments}
V.I.M. was partially supported by the Russian Foundation for Basic Research under Projects Nos. 07-
02-00598 and 09-02-00142.

\end{document}